\documentclass[runningheads]{llncs}
\usepackage[T1]{fontenc}
\usepackage[utf8]{inputenc}
\usepackage{kpfonts}
\usepackage{booktabs}
\usepackage{siunitx}
\usepackage{caption}
\usepackage{microtype}
\usepackage{varioref}
\usepackage[autostyle]{csquotes}
\usepackage{subcaption}
\usepackage{mathdots} 
\usepackage[backend=biber,safeinputenc,sorting=none]{biblatex}
\usepackage{hyperref}
\usepackage{physics}
\usepackage{braket}
\usepackage[final]{pdfpages}
\usepackage{tikz}
\usetikzlibrary{quantikz}
\usepackage{comment}
\usepackage{leftindex}
\usepackage[toc,page]{appendix}
\usepackage{graphicx}
\begin{document}
\title{Quantum algorithm for anisotropic diffusion and convection equations with vector norm scaling}
\titlerunning{Quantum algorithm for anisotropic diffusion and convection equations}
\author{Julien Zylberman\inst{1}%\orcidID{0000-0002-7785-8246} 
\and
Thibault Fredon\inst{2}%\orcidID{0009-0005-9427-9309} 
\and
Nuno F.Loureiro\inst{2}%\orcidID{0000-0001-9755-6563}
\and Fabrice Debbasch \inst{1} %\orcidID{0000-0001-6894-6159}
}
\authorrunning{J.Zylberman et al.}
\institute{Sorbonne Université, Observatoire de Paris, Université PSL, CNRS, LUX, F-75005 Paris, France \and
Plasma Science and Fusion Center, Massachusetts Institute of Technology, Cambridge, Massachusetts 02139, USA}
%\email{}
%
\maketitle              % typeset the header of the contribution

\begin{abstract}

In this work, we tackle the resolution of partial differential equations (PDEs) on digital quantum computers. Two fundamental PDEs are addressed: the anisotropic diffusion equation and the anisotropic convection equation. We present a quantum numerical scheme consisting of three steps: quantum state preparation, evolution with diagonal operators, and measurement of observables of interest. The evolution step relies on a high-order centered finite difference and a product formula approximation, also known as Trotterization. We provide novel vector-norm analysis to bound the different sources of error. We prove that the number of time-steps required in the evolution can be reduced by a factor $\Theta (16^n)$ for the diffusion equation, and $\Theta (4^n)$ for the convection equation, where $n$ is the number of qubits per dimension, an exponential reduction compared to the previously established operator-norm analysis.
\end{abstract}

\section{Introduction}

The development of quantum numerical schemes for solving linear and non-linear partial differential equations is an active field of research, with open question about the computational efficiency, the reachable accuracies and the potential speedups when compared to classical methods. While the simulation of quantum systems by quantum computers has been extensively studied \cite{ kassal2008polynomial,kivlichan2017bounding,babbush2017exponentially,low2017optimal,low2019hamiltonian,babbush2019quantum}, the simulation of classical (non-quantum) systems requires the development of novel quantum numerical schemes. Those need to take into account non-quantum evolution, i.e., non-unitary or non-linear evolution, that are not directly implementable on quantum computers.

In the following, we address the problem of solving anisotropic convection and anisotropic diffusion equations on quantum computers. First, we introduce a quantum numerical scheme based on high-order central finite difference and product formula approximation, also known as Trotter-splitting. Then, we present a numerical analysis based on vector norm instead of generic operator norm analysis\footnote{In the literature, quantum algorithms for simulating Hamiltonian systems, or solving differential equations, often have a complexity given in terms of the spectral norm of the associated operator, such as the Hamiltonian or the operator $\hat{A}$ when solving a linear ordinary differential equation $\partial_tf=\hat{A}f$.}. We show that the vector norm analysis predicts a reduction of the number of time-steps by an exponential factor $O(4^n)$ for the anisotropic convection equation, and by a factor $O(16^n)$ for the anisotropic diffusion equation, where $n$ is the number of qubits per dimension.

\section{Problem and approach}

The first considered PDE is the $d$-dimensional anisotropic diffusion equation: \begin{equation}
    \partial_t\phi=\vec{\nabla}\cdot(K(\vec{x},t)\vec{\nabla}\phi),
\label{eq:diffusion eq}
\end{equation} 
where $\phi$ is the unknown, $\vec{\nabla}=(\partial_{x_1},\hdots,\partial_{x_d})$ and $K(\vec{x},t)=\text{diag}(\kappa_1(\vec{x},t),\hdots,\kappa_d(\vec{x},t))$ is a space- and time-dependent thermal conductivity. The second PDE of interest is the $d$-dimensional anisotropic convection equation:
\begin{equation}
    \partial_t f +\vec{c}(\vec{x},t).\vec{\nabla}f=0,
\label{eq:convection eq}
\end{equation}
where $\vec{c}(\vec{x},t)=(c_1(\vec{x},t),\hdots,c_d(\vec{x},t))$ is a space- and time-dependent velocity field. Both equations are defined on the $d$-dimensional torus $\mathbb{T}^d$, i.e., the box $[0,1]^d$ with periodic boundary conditions, with a time $t\in[0,T]$ for some $T>0$, and with initial conditions $\phi(\vec{x},t=0)=\phi_0(\vec{x})$ and $f(\vec{x},t=0)=f_0(\vec{x})$ respectively. Additionally, we assume that the $c_j$ and the $\kappa_j$ functions are $2p+1$ differentiable, for some integer $p$, and do not depend on the $j$-th variables: $\forall j \in \{1,\hdots, d\}$,  $\partial_{x_j}c_j=0$, $\partial_{x_j}\kappa_j=0$\footnote{This assumption will be necessary to derive an efficient quantum circuit made only of quantum Fourier transforms and diagonal operators for the evolution step.}.

\paragraph{Real-space encoding.}
In order to numerically solve these equations on quantum computers, we defined the real space encoding of the solution function into a $n=n_1+...+n_d$ qubit state as: $\ket{f}_t=\sum_{\vec{x} }f(\vec{x},t)\ket{\vec{x}} \in \mathcal{H}$, where $\mathcal{H}=\mathcal{H}_{x_1}\otimes...\otimes\mathcal{H}_{x_d}$ is the Hilbert space associated to the $n$ entangled qubits and each $\mathcal{H}_{x_j}$ is a $2^{n_j}$-dimensional Hilbert space. The ket vector $\ket{\vec{x}}$ is defined as the tensor product of the ket vector of each space $\ket{\vec{x}}=\ket{x_1}\otimes...\otimes\ket{x_d}$. Each ket vector $\ket{x_j}$ is defined as $\ket{x_j}=\ket{q_0}\otimes...\otimes\ket{q_{n_j-1}}$, with $x_j=\sum_{k=0}^{n_j-1}q_k/2^{k+1}\in  \{0,\hdots,2^{n_j}-1\}$ being the dyadic expansion of $x$ and $\forall k, q_k\in\{0,1\}$.

\paragraph{Quantum state preparation.} The quantum numerical scheme is based on three steps. The first one is a quantum state preparation routine that encodes the initial condition into an $n$-qubit state: $    \phi_0\rightarrow\ket{\phi_0}=\frac{1}{\mathcal{N}}\sum_{\vec{x}}\phi_0(\vec{x})\ket{\vec{x}}$,   $f_0\rightarrow\ket{f_0}=\frac{1}{\mathcal{N}'}\sum_{\vec{x}}f_0(\vec{x})\ket{\vec{x}}$,
 where $\mathcal{N},\mathcal{N}'$ are normalization constants. Many quantum state preparation routines have been published in the litterature: exact protocols generically require a number of quantum gates proportionate to the number of components of the target quantum states. In other word, optimal circuit size\footnote{The size of a quantum circuit is the number of primitive quantum gates.} for preparing arbitrary $n$-qubit state is $O(2^n)$  \cite{zhang2022quantum,yuan2023optimal,sun2023asymptotically}. Fortunately, efficient methods\footnote{A quantum circuit is said efficient is its circuit complexity expressed in terms of size, depth and ancilla qubits that scale polynomially with $n$ and $1/\epsilon$.} have been developed for qubit states depending on differentiable functions, which leverage the structure of the state by using Walsh, Fourier or polynomial series \cite{zylberman2024efficient,zylberman2025efficient,moosa2023linear,rosenkranz2025quantum}.
 
\begin{figure}
\centering
\begin{quantikz}
\lstick[wires=1]{$\ket{x_1}$}
&\gate[5,nwires={3,4},style={rounded
corners},disable auto height]{\begin{matrix}
\text{Quantum State Preparation:}\\ \text{}\\\text{Walsh Series Loader}\\\text{Fourier Series Loader}\\\text{Polynomial Series Loader}\\\text{...}\end{matrix}} \slice{$\ket{f_0}$ or $\ket{\phi_0}$} &\gate[5,nwires={3,4},style={rounded
corners},disable auto height]{\begin{matrix}
\text{Evolution:}\\\text{}\\\text{Diagonal operations and QFTs}\end{matrix}} \slice{$\ket{f}_T$ or $\ket{\phi}_T$} &\gate[5,nwires={3,4},style={rounded
corners},disable auto
height]{\begin{matrix}
\text{Measurement protocol:}\\ \text{}\\\text{Hadamard test}\\\text{Swap test}\\\text{Quantum Amplitude Estimation}\\ \\\rightarrow \leftindex_T{\bra{f}}\hat{O}\ket{f}_T \text{ or } \leftindex_T{\bra{\phi}}\hat{O}\ket{\phi}_T \end{matrix}} &\\
 \lstick[wires=1]{$\ket{x_2}$}& \qw & \qw & \qw & \\
 &&&&&&&&&&&\\
  & && && && &&&&& \\
 \lstick[wires=1]{$\ket{x_d}$} &\qw&\qw &\qw& & 
\end{quantikz}
\caption{Quantum numerical scheme associated with the resolution of the $d$-dimensional convection or diffusion equations with position- and time- dependent coefficients.}
\label{fig:quantum numerical scheme}
\end{figure}
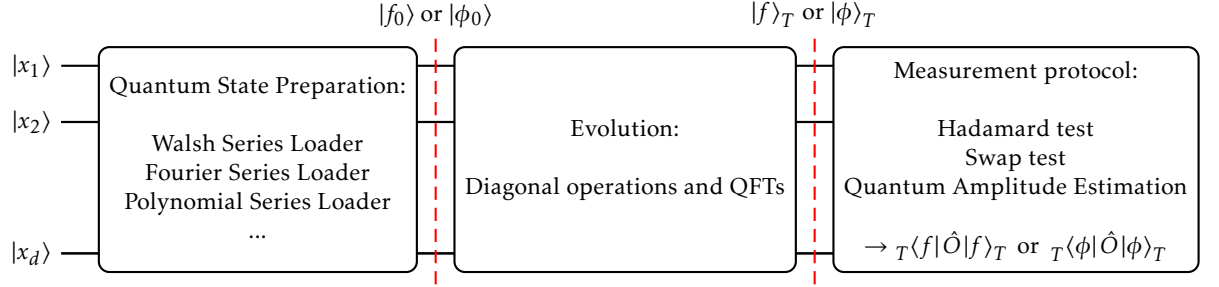

\paragraph{Evolution.} The second step reproduces the PDE evolution using primitive quantum gates,  evolving the initial qubit state $\ket{\phi_0}$ (resp. $\ket{f_0}$) toward a qubit state that encodes the PDE's solution at a target time $T>0$: $\ket{\phi_0}\rightarrow \ket{\phi}_T= \frac{1}{\mathcal{N}''}\sum_{\vec{x}}\phi(\vec{x},T)\ket{\vec{x}}$  (resp. $\ket{f_0}\rightarrow \ket{f}_T= \frac{1}{\mathcal{N}''}\sum_{\vec{x}}f(\vec{x},T)\ket{\vec{x}}$). The evolution step is based on a high-order centered finite difference $\partial_{x_i}\phi(\vec{x},t) \rightarrow  (1/\Delta x_i)\sum_{k=-p}^p a_k\phi(\vec{x}+k\vec{\Delta x_i},t)+O(\Delta x_i^{2p})$, where  $a_0=0$ and, for $k\in \{-p,-p+1,...p-1,p\}\backslash \{0\}$, $a_k=(-1)^{k+1}(p!)^2/(k(p-k)!(p+k)!)$\footnote{The $a_k$ coefficients are the finite difference coefficients, solution of the system of equations $\sum_{k=-p}^p a_k k^j=\delta_{j,1}$}.
The finite difference recasts the PDE problems as ordinary differential equation (ODE) problems:
\begin{equation}
    \partial_t \ket{\tilde{f}}_t =-i(\sum_{j=1}^d \hat{c}_j(t)\hat{D}_j)\ket{\tilde{f}}_t
\label{eq:ODE convection}
\end{equation}
\begin{equation}
    \partial_t \ket{\tilde{\phi}}_t =-(\sum_{j=1}^d \hat{\kappa}_j(t)\hat{D}_j^2)\ket{\tilde{\phi}}_t,
\label{eq:ODE diffusion}
\end{equation}
with the initial conditions $    \ket{\tilde{\phi}}_{t=0}=\ket{\phi_0}$, $    \ket{\tilde{f}}_{t=0}=\ket{f_0}$ and where $\hat{D}_{j}=\hat{I}_1\otimes\cdots\otimes \hat{I}_{j-1} \otimes -i\frac{\sum_{k=-p}^p a_k (\hat{S}_j)^k}{\Delta x_j}\otimes \hat{I}_{j+1} \otimes\cdots \otimes \hat{I}_d$ is the discrete derivative operator associated with the $j$-th axis and $\hat{c}_j(t)=c_j(\hat{\vec{X}},t)$, $\hat{\kappa}_j(t)=\kappa_j(\hat{\vec{X}},t)$ with  $\hat{\vec{X}}=(\hat{X}_1,...\hat{X}_d)$, $\hat{X}_j=\hat{I}_1\otimes \cdots \otimes \hat{I}_{j-1}\otimes \sum_{X_j}X_j\ket{X_j}\bra{X_j}\otimes \hat{I}_{j+1}\otimes \cdots\otimes \hat{I}_d$.

The solutions of the ODEs are given by  time-ordered evolution operators as: $\ket{\tilde{f}}_t=\mathcal{T}e^{-i\int_0^t\sum_{j=1}^d \hat{c}_j(s)\hat{D}_jds}\ket{f_0}$ and $\ket{\tilde{\phi}}_t=\mathcal{T}e^{-\int_0^t\sum_{j=1}^d \hat{\kappa}_j(s)\hat{D}_j^2ds}\ket{\phi_0}$, where $\mathcal{T}$ is the time-ordering operator. In general, it is not possible to implement directly time-ordered exponentials on computers, whether classical or quantum. To overcome this issue, we consider first order product formula to decompose the time-ordered evolution operators into efficiently implementable unitary operators. The standard product formula denoted by the index $s$ and the generalized one denoted by $g$ are defined as \cite{An2021}:
\begin{equation}
\begin{split}
    \hat{U}_{s}(t+h,t)=\overleftarrow{\prod_{j=1}^{d}} \exp(-ih \hat{c}_j(t+h)\hat{D}_j), \text{ } \text{ } \text{ } \text{ }
    \hat{U}_{g}(t+h,t)=\overleftarrow{\prod_{j=1}^{d}} \exp(-i\int_{t}^{t+h}\hat{c}_j(s)ds\hat{D}_j) \\
    \hat{V}_{s}(t+h,t)=\overleftarrow{\prod_{j=1}^{d}} \exp(-h \hat{\kappa}_j(t+h)\hat{D}_j^2), \text{ } \text{ } \text{ } \text{ }
    \hat{V}_{g}(t+h,t)=\overleftarrow{\prod_{j=1}^{d}} \exp(-\int_{t}^{t+h}\hat{\kappa}_j(s)ds\hat{D}_j^2) 
\label{product formula}
\end{split}
\end{equation}
where $\overleftarrow{\prod_{j=1}^d}A_j=A_d...A_1$ is a product with decreasing indexes for non-commuting operators.  The product formula approximation is as accurate as $h$ is small. Therefore, we discretize the time span $[0,T]$ into $L$ equal segments and we define $\ket{\tilde{\phi}_\alpha}_t$ and $\ket{\tilde{f}_\alpha}_t$, with $\alpha=s,g$, the qubit states given by $L$  trotterized-time-steps as:
\begin{equation}
\begin{split}
\label{eq:L time step evolution}
  \ket{\tilde{\phi}_{\alpha}}_T=\prod_{l=1}^L\hat{V}_{\alpha}(\frac{lT}{L},\frac{(l-1)T}{L})\ket{\phi_0}
   \\ \ket{\tilde{f}_{\alpha}}_T=\prod_{l=1}^L\hat{U}_{\alpha}(\frac{lT}{L},\frac{(l-1)T}{L})\ket{f_0}
\end{split}
\end{equation}

The implementation of the operators $\hat{V}_\alpha$ and $\hat{U}_\alpha$ is performed using quantum Fourier transforms and diagonal operators. The quantum Fourier transform (QFT) diagonalizes the discrete derivative operators $\hat{D}_j$ as  $\hat{D}_j= \widehat{QFT}_j^{-1}(\sum_{X_j}d(X_j)\newline \ket{X_j}\bra{X_j}) \widehat{QFT}_j$,
where $\widehat{QFT}_j$ is the QFT acting on the $j$-th register and $d(X_j)=\frac{2}{\Delta x_j}\sum_{q=0}^p a_q\sin(2\pi q X_j)$ is one of the eigenvalues of the operator $\hat{D}_j$. Consequently, the operators $\hat{D}_j^2$ and each of the exponential defined in Equation \ref{product formula} are also diagonalized by the QFT.
The associated quantum circuit is well established for the QFT \cite{nielsen2010quantum}. The diagonal operators depends on differentiable functions and can be efficiently implemented using (sparse) Walsh or Fourier series approximations \cite{welch2014efficient,zylberman2025efficient}.

\paragraph{Measurement protocol.} The last step of the numerical scheme is the extraction of information from the final qubit state $\ket{\phi}_T$ (resp. $\ket{f}_T$) using measurement protocols such as quantum amplitude estimation, the Hadamard test or the Swap test \cite{nielsen2010quantum,brassard2000quantum,grinko2021iterative,rall2023amplitude,giurgica2022low,shang2024estimating}. These protocols enable the measurement of averaged values of observables $\ket{\phi}_T\rightarrow \leftindex_T{\bra{\phi}}\hat{O}\ket{\phi}_T$, where $\hat{O}$ is designed to correspond to relevant information, such as the value of the solution at a given position, the average position of the solution, its standard deviation, or higher-order moments. 

\section{Error analysis}
\paragraph{Space discretization error.}

The quantum numerical scheme introduced two approximations: the space-discretization error and the product formula error.
The first one is quantified by the difference between $\ket{f}_t$, the normalized vector encoding the solution of the convection equation Eq.(\ref{eq:convection eq}) at time $t$, and $\ket{\tilde{f}}_t$, the solution of the ODE Eq.(\ref{eq:ODE convection}). One can bound their difference by writing the following equation for $\ket{f}_t$: $\partial_t \ket{f}_t=-i\sum_{j=1}^d \hat{c}_j(t)\hat{D}_j\ket{f}_t+\ket{r(t)}$ with $\ket{r(t)}= \sum_{j=1}^d\hat{c}_j(t)(i\hat{D}_j\ket{f}_t-\ket{\partial_{x_j}f}_t)$. The solution of this equation is given by variation of parameter formula as $ \ket{f}_t=\hat{U}(t,0)\ket{f_0}+\int_0^t \hat{U}(t,s)\ket{r(s)}ds$, where $\hat{U}$ is the time-ordered evolution operator associated with the ODE Eq.(\ref{eq:ODE convection}). The variation of parameter formula makes it possible to bound the difference as $||\ket{f}_t-\ket{\tilde{f}}_t||_{2,N}\leq t \max_{s\in[0,t]}\|\ket{r(s)}\|_{2,N}$. Then, using Taylor inequality, one can show $i\hat{D}_j\ket{f}=\ket{\partial_{x_j} f}+O((\Delta x_j)^{2p})$. Finally, the difference is bounded as
$\|\ket{\tilde{f}}_T-\ket{f}_T\|_{2,N} \leq t K \sum_{j=1}^d ||c_j||_\infty (\Delta x_j)^{2p} $, where $K$ is a constant depending on $p$ and the maximum value over $j\in\{0,...,d\}$ and time of the $(2p+1)$ derivative of $f_0$ with respect to axis $j$. Since $\Delta x=1/2^n$, one can choose a number of qubits associated with the $j$-th axis as $n_j=\lceil\frac{1}{2p}\log_2(TK\|c_j\|_{\infty}/\epsilon)\rceil$ to ensure an $\epsilon$ approximation level.

One can proceed analogously to bound the discretization error associated with the diffusion equation. Since the diffusion equation has a non-unitary evolution\footnote{The evolution operator $\hat{U}$ associated with the discretized convection equation Eq.(\ref{eq:ODE convection}) is unitary thanks to the central finite difference and the assumption that $\forall j, \partial_{x_j}c_j=0$, while the evolution operator associated with the discretized diffusion equation Eq.(\ref{eq:ODE diffusion}) is non-unitary, but has a spectral norm $\|\hat{V}\|\leq 1$.}, one needs to consider the difference between the normalized vectors as: $\|\frac{\ket{\tilde{\phi}}_t}{\|\ket{\tilde{\phi}}_t\|_{2,N}}-\frac{\ket{\phi}_t}{\|\ket{\phi}_t\|_{2,N}}\|_{2,N}\leq TK\sum_{j=1}^d\|\kappa_j\|_{\infty}(\Delta x_j)^{2p}$,
where $K$ is independent of $N$ and depends on the the $2p+1$ derivatives of $\phi$.

\paragraph{Product formula approximation.}

%Using the variation of parameter formula and Taylor's theorem, one can prove that the difference between $\ket{\tilde{f}_{\alpha}}_T$ and $\ket{\tilde{f}}_T$ scales as $O(T^2/L)$. The number of time steps $L$ to reach an arbitrary precision $\epsilon>0$ should therefore scale as $L=O(T^2/\epsilon)$. One can prove that the prefactor in $L$ do not scale with $n$. 

A direct expansion of the Trotter formula gives $\|\big(e^{i(\hat{H}_1+\hat{H}_2)\Delta t}-e^{i\hat{H}_1 \Delta t}e^{i\hat{H}_2 \Delta t}\big)\ket{\psi}\|_{2,N}= \frac{\Delta t^2}{2}\|[\hat{H}_2,\hat{H}_1]\ket{\psi}\|_{2,N}+O(\Delta t^3)$. In the literature on quantum algorithms, most of the scaling uses the operator norm inequality by bounding the commutator as $\|[\hat{H}_2,\hat{H}_1]\ket{\psi}\|_{2,N}\leq \|[\hat{H}_2,\hat{H}_1]\|_{2}\leq 2\|\hat{H}_1\|_{2}\|\hat{H}_2\|_{2}$ \footnote{The first reference introducing vector norm analysis for quantum algorithm is \cite{An2021}, which address the problem of solving the Schrödinger equation with a bounded potential, i.e., in the case where $\hat{H}_1$ is asymptotically unbounded and $\hat{H}_2$ is bounded. Our results generalize their approach since all the summands are composed of discrete derivative operators that are asymptotically unbounded for the convection and diffusion equations.}. However, in the case of the convection or the diffusion equation, the summand $\hat{H}_j$ depends on the discrete derivative operator such as $\|\hat{H}_j\|_{2}=\|c_j\hat{D}_j\|_2=O(1/\Delta x_j)=O(2^{n_j})$ or $\|\hat{H}_j\|_{2}=\|\kappa_j\hat{D}_j^2\|_2=O(1/\Delta x_j^2)=O(4^{n_j})$, resulting in exponential-with-$n$ error bounds. In the following, we overcome these exponential scaling by considering the action of the operator $\hat{D}_j$ on the qubit state encoding the solution $\ket{f}_t$ (or $\ket{\phi}_t$), deriving a novel bound in vector norm with terms of the form $\|[\hat{H}_2,\hat{H}_1]\ket{f}_t\|_{2,N}=O(1)$. Precisely, one can prove that for the convection equation and $\alpha=s,g$:
\begin{equation}
\begin{split}
    ||\ket{\tilde{f}_{\alpha}}_T-\ket{\tilde{f}}_T||&\leq  a_\alpha \frac{T^2}{L}+ r_\alpha \\
\end{split}
\label{Eq: Scaling}
\end{equation}
where $r_\alpha$ contains higher order terms that are asymptotically negligible in the large $L \rightarrow +\infty$ limit and

\begin{equation*}
\begin{split}    
a_g&=\frac{1}{2}  \sum_{j=1}^{d-1} \sum_{m=j+1}^d \bigg( \|c_j\|_{\infty} \|\partial_{x_j} c_m\|_{\infty} \max_{t\in[0,T]}\frac{ \|\partial_{x_m}f_t\|_{2,N}}{\|f_0\|_{2,N}}
     +\|c_m\|_{\infty} \|\partial_{x_m} c_j\|_{\infty} \max_{t\in[0,T]}\frac{ \|\partial_{x_j}f_t\|_{2,N}}{\|f_0\|_{2,N}} \bigg) \\ %& \leq\frac{d(d-1)}{2}\max_{1\leq m<j\leq d, t\in[0,T]} ||c_j||_{\infty}||\partial_{x_j}c_m||_{\infty}\frac{||\partial_{x_m}f_t||_{2,N}}{||f_0||_{2,N}} \\
a_s&=a_g +\frac{1}{2}\sum_{j=1}^d\|\partial_tc_j\|_{\infty} \max_{t\in[0,T]}\frac{\|\partial_{x_j}f_t\|_{2,N}}{\|f_0\|_{2,N}} \leq a_g+ \frac{d}{2}\max_{1\leq j \leq d, t\in [0,T]} ||\partial_tc_j||_{\infty} \frac{||\partial_{x_j} f_t||_{2,N}}{||f_0||_{2,N}}
\end{split}
\end{equation*}
Remark that $||\partial_{x_m}f_t||_{2,N}/||f_0||_{2,N}$ converges toward  $||\partial_{x_m}f_t||_{L^2}/||f_0||_{L^2}=||\partial_{x_m}f_0||_{L^2}/||f_0||_{L^2}$ in the large $N$ limits, which depends only on $f_0$\footnote{The $L^2$ norm of any derivative of $f$ is constant over time, equal to its initial value at time $t=0$.}.

Similarly, for the diffusion equation, one can show:
\begin{equation}
   \left  \|\frac{\ket{\tilde{\phi}_\alpha}_T}{\|\ket{\tilde{\phi}_\alpha}_T\|_{2,N}}-\frac{\ket{\tilde{\phi}}_T}{\|\ket{\tilde{\phi}}_T\|_{2,N}}\right\|_{2,N}\leq a_\alpha' \frac{T^2}{L}+r_\alpha'
\end{equation}

where $r_\alpha'$ contains higher order terms that are asymptotically negligible and

\begin{equation*}
\begin{split}    
a_g'&=\frac{1}{2}  \sum_{j=1}^{d-1} \sum_{m=j+1}^d \bigg( \|\kappa_j\|_{\infty} \|\partial_{x_j}^2 \kappa_m\|_{\infty} \max_{t\in[0,T]}\frac{ \|\partial_{x_m}^2\phi_t\|_{2,N}}{\|\tilde{\phi}_T\|_{2,N}}
     +\|\kappa_m\|_{\infty} \|\partial_{x_m}^2 \kappa_j\|_{\infty} \max_{t\in[0,T]}\frac{ \|\partial_{x_j}^2\phi_t\|_{2,N}}{\|\tilde{\phi}_T\|_{2,N}} \bigg) \\ %& \leq\frac{d(d-1)}{2}\max_{1\leq m<j\leq d, t\in[0,T]} ||\kappa_j||_{\infty}||\partial_{x_j}^2\kappa_m||_{\infty}\frac{||\partial_{x_m}^2\phi_t||_{2,N}}{||\tilde{\phi}_T||_{2,N}} \\
a_s'&=a_g' +\frac{1}{2}\sum_{j=1}^d\|\partial_t\kappa_j\|_{\infty} \max_{t\in[0,T]}\frac{\|\partial_{x_j}^2\phi_t\|_{2,N}}{\| \tilde{\phi}_T\|_{2,N}} \leq a_g+ \frac{d}{2}\max_{1\leq j \leq d, t\in [0,T]} ||\partial_t\kappa_j||_{\infty} \frac{||\partial_{x_j}^2 \phi_t||_{2,N}}{||\tilde{\phi}_T||_{2,N}}
\end{split}
\end{equation*}
where $||\tilde{\phi}_T||_{2,N}=||\widehat{QFT}\tilde{\phi}_T||_{2,N}\ge |\bra{0}^{\otimes n}\widehat{QFT}\ket{\phi}_T|=|\sum_{\vec{X}}\tilde{\phi}(\vec{X},T)|/\sqrt{N}=|\sum_{\vec{X}}\phi_0(\vec{X})|/\sqrt{N}$, where the average value of $\tilde{\phi}$ is conserved over time and equal to the average value of $\phi_0$. Then, by assuming that the average value of $\phi_0$ is non-zero, one deduce that $1/||\tilde{\phi}_T||_{2,N}\leq \sqrt{N}/(|\sum_{\vec{X}}\phi_0(\vec{X})|)\xrightarrow[]{N\rightarrow +\infty} 1/(\sqrt{N}\braket{\phi_0}_{\mathbb{T}^d})$ with $\braket{\phi_{0}}_{\mathbb{T}^d}=\int_{\mathbb{T}^d}\phi_{0}(\vec{x})d\vec{x}$ and $\|\partial_{x_m}^2\phi_t\|_{2,N}/\|\tilde{\phi}_T\|_{2,N} \leq$ $\sqrt{N}\|\partial_{x_m}^2\phi_t\|_{2,N}/|\sum_{\vec{X}}\phi_0(\vec{X})|\xrightarrow[]{N\rightarrow +\infty}||\partial_{x_m}\phi_t||_{L^2}/\braket{\phi_0}_{\mathbb{T}^d}$.

In both case, the product formula approximation scales as $O(T^2/L)$, with a prefactor independent of the discretization step. Consequently, the number of time steps $L$ to reach an arbitrary precision $\epsilon>0$ should scale as $L=O(T^2/\epsilon)$. It represents an exponential-with-$n$ reduction compared to operator norm analysis that predicts a number of time-step scaling with the operator norms $\|\hat{D}_j^2\|_2=O(4^{n_j})$ for the convection equation, and $\|\hat{D}_j^4\|_2=O(16^{n_j})$ for the diffusion equation.

%A novel numerical analysis for bounding the various sources of error is presented. The discretization error is bounded by a term proportional to $(\Delta x)^{2p}$. For the product formula approximation, the error is typically measured using the operator norm of the evolution operator. This error depends on the space- and time-dependent coefficients $\vec{c}$ or $K$, and on the operator norm of the discrete derivative operators. The latter are asymptotically unbounded, meaning their spectral norm scales proportionally to $1/\Delta x=2^n$, where $n$ is the number of qubits. As a result, the number of time steps required to guarantee a target accuracy scales exponential with $n$. However, the operator norm is a worst-case scenario, whereas in practice, the initial condition possesses certain regularity properties. We prove that, under appropriate assumptions on the initial condition and the position- and time-dependent coefficients, the product formula approximation does not scale with the operator norm of the discrete derivative operators. We present a vector norm analysis that guarantees similar accuracy with exponentially fewer time steps than analyses based on the operator norm.
%The numerical analysis allows to derive the query complexity of the quantum numerical scheme by estimating the number of quantum state preparation, diagonal operations and operator $\hat{O}$ required to extract relevant information.

\section{Conclusion}
In this work, we introduced quantum numerical schemes for an anisotropic convection equation and for an anisotropic diffusion equation. The numerical schemes are composed of three steps: a quantum state preparation to load the initial condition into a qubit state; an evolution step made of diagonal operators and quantum Fourier transforms; and a measurement protocol to estimate quantities of interest from the numerical solution. We presented an error analysis of the space discretization and product formula approximations. In particular, we introduced novel computations based on vector norm analysis to prove that the required number of time-steps needed to reach a given accuracy $\epsilon>0$ can be reduced by an exponential-with-$n$ factor, compared to operator norm analysis. These results pave the way for the development  of efficient quantum numerical schemes for other PDE problems such as the Vlasov equation or the Fokker-Planck equation and for the development of novel numerical analysis for other types of quantum numerical schemes.

\section{Acknowledgement}
This preprint has not undergone peer review or any post-submission improvements or corrections. The Version of Record of this contribution is published in \textit{Quantum Engineering Sciences and Technologies for Industry and Services} (QUEST-IS 2025), and is available online at \url{https://doi.org/10.1007/978-3-032-13855-2_23}.

\printbibliography

%
%
%=
\end{document}